\documentclass[%
 reprint,
superscriptaddress,
showpacs,preprintnumbers,
nofootinbib,
 amsmath,amssymb, 
 aps,
 prd,
 longbibliography,
]{revtex4-1}

\usepackage{cancel}
\usepackage{accents}
\usepackage{mciteplus,slashed}
\usepackage{amssymb,cancel,amsmath}
\usepackage{dcolumn}
\usepackage{bm}
\usepackage[caption=false]{subfig}
\usepackage{appendix}
\usepackage{physics}
\usepackage{booktabs}
\usepackage{feynmp-auto}
\usepackage[T1]{fontenc}	
\usepackage{csvsimple}
\usepackage[bookmarksnumbered=true]{hyperref} 
\hypersetup{
    colorlinks = true,
    linkcolor = blue,
    anchorcolor = blue,
    citecolor = blue,
    filecolor = blue,
    urlcolor = blue
    }
\usepackage[section]{placeins}
\usepackage[capitalise]{cleveref}
\usepackage{booktabs}
\usepackage{graphicx}
\usepackage{mathrsfs}
\usepackage[caption=false]{subfig}
\graphicspath{{Figures/}}

\AtBeginDocument{
\heavyrulewidth=.08em
\lightrulewidth=.05em
\cmidrulewidth=.03em
\belowrulesep=.65ex
\belowbottomsep=0pt
\aboverulesep=.4ex
\abovetopsep=0pt
\cmidrulesep=\doublerulesep
\cmidrulekern=.5em
\defaultaddspace=.5em
}
\usepackage[dvipsnames]{xcolor}
\usepackage[normalem]{ulem}
\usepackage{fontawesome} 

\def\e{\mathrm{e}}

\graphicspath{{Figures/}}

\begin{document}

\title{Long-lived particles and the Quiet Sun }

\preprint{CALT-TH/2023-023}

\author{R. Andrew Gustafson}
\email{gustafr@vt.edu}
\affiliation{Center for Neutrino Physics, Department of Physics,
Virginia Tech, Blacksburg, VA 24061, USA}

\author{Ryan Plestid}
\email{rplestid@caltech.edu}
\affiliation{Walter Burke Institute for Theoretical Physics, California Institute of Technology, Pasadena, CA 91125, USA}

\author{Ian M. Shoemaker}
\email{shoemaker@vt.edu}
\affiliation{Center for Neutrino Physics, Department of Physics,
Virginia Tech University, Blacksburg, VA 24061, USA}

\author{Albert Zhou}
\affiliation{Institut f\"ur Astroteilchenphysik, Karlsruher Institut f\"ur Technologie (KIT), D-76021 Karlsruhe, Germany}
 
\begin{abstract} 
     The nuclear reaction network within the interior of the Sun is an efficient MeV physics factory, and can produce long-lived particles generic to dark sector models. In this work we consider the sensitivity of satellite instruments, primarily the RHESSI Spectrometer, that observe the Quiet Sun in the MeV regime where backgrounds are low. We find that Quiet Sun observations offer a powerful and complementary probe in regions of parameter space where the long-lived particle decay length is longer than the radius of the Sun, and shorter than the distance between the Sun and Earth. We comment on connections to recent model-building work on heavy neutral leptons coupled to neutrinos and high-quality axions from mirror symmetries. 
\end{abstract}

\maketitle 
 
\section{Introduction \label{Introduction}}
It has long been recognized that the solar interior can serve as an efficient factory for keV-scale physics beyond the Standard Model (BSM), e.g.\ solar axions and dark photons \cite{Sikivie:1983ip,Dimopoulos:1986kc,Avignone:1986vm,Raffelt:1990yz,Battesti:2007um,Pospelov:2008jk,An:2014twa,Graham:2015ouw}. In addition to thermal production mechanisms, nuclear reactions within the Sun may also source BSM particles up to masses and energies of roughly $ 15~{\rm MeV}$ \cite{Raffelt:1982dr,Raffelt:1990yz,CAST:2009klq,Borexino:2012guz,Borexino:2013bot,Plestid:2020ssy,Plestid:2020vqf}. If a flux of long-lived particles (LLPs) in this energy regime emanates from the solar interior, they may transit toward the Earth and their decay products can leave detectable signatures. It is important to emphasize that LLPs are generic consequences of a dark sector with relatively light particles and feeble couplings to the SM \cite{Beacham:2019nyx,Agrawal:2021dbo,Curtin:2018mvb,Knapen:2022afb}. As decay lengths become long, LLPs become increasingly difficult to detect and strategies to attack this ``lifetime frontier'' are valuable tools in the search for BSM physics. This idea has been previously investigated, largely considering FERMI-LAT, in the high energy, i.e.\ $\gtrsim 100~{\rm MeV}$, regime for annihilating dark matter \cite{Schuster:2009au,Batell:2009zp,PhysRevD.95.123016,Leane:2021tjj}.

In this work we point out that existing data from the RHESSI satellite spectrometer \cite{smith2003rhessi}, which observed the Quiet Sun,\!\footnote{Time periods without intense surface activity such as solar flares.} can place interesting limits on dark sectors with LLPs in the range of $\mathcal{O}(100\, \rm keV) - \mathcal{O}(1 \, \rm MeV)$. This is an old idea, first proposed by Stodolsky and Raffelt in 1982 in the context of a 200~{\rm keV} axion \cite{Raffelt:1982dr}, however, it has remained unexplored despite new data in the intervening decades \cite{Dennis:2022mso}. We illustrate the potential sensitivity of Quiet Sun data with a number of BSM examples, emphasizing different production mechanisms which may operate in this mass window. A conservative analysis of existing data from RHESSI is capable of offering complimentary constraints on production mechanisms involving neutrino upscattering, and can probe previously untouched regions of parameter space for axion like particles (ALPs) with masses close to $\sim 1~ {\rm MeV}$. Upcoming missions, such as the COSI satellite \cite{COSI_2023,Tomsick:2021wed}, may be able to substantially improve on the capabilities of RHESSI by {\it i)} taking advantage of a larger instrument surface area, {\it ii)} making use of dead time to carefully study backgrounds, and {\it iii)} taking advantage of distinctive spectral features.

We focus on LLPs that decay primarily to photons,\!\footnote{We could also consider decays to $e^+e^-$ pairs but an analysis is complicated by the magnetic fields that surround the Earth.} and have decay lengths, $\ell_{\rm LLP}$, that satisfy 
\begin{equation}
    R_{\odot} \ll \ell_{\rm LLP} \ll d_{\odot}~,
    \label{eq:dec_range}
\end{equation}
where $R_{\odot}$ is the radius of the Sun and $d_{\odot}$ is the distance from the Sun to the Earth. This allows an $O(1)$ fraction of the LLPs to decay en-route to the satellite instrument. In this limit, the flux of LLPs will never reach any terrestrial experiment since they will decay in flight and their daughter photons will be absorbed in the upper atmosphere. In this sense, Quiet Sun observations are complimentary to terrestrial searches for LLPs from the Sun such as those that have been performed by CAST \cite{CAST:2009klq} and Borexino \cite{Borexino:2012guz}. 

We perform a straightforward (and conservative) rate-only analysis, the details of which can be found at the end of \cref{sec:Neutrinos}. In the body of the paper we organize our discussion along the lines of specific BSM scenarios. We discuss neutrino upscattering in \cref{sec:Neutrinos} and solar axion production in \cref{sec:Axions}. We also spend time focusing on model-independent LLP constraints in \cref{sec:Model_Ind}. In \cref{Future} we discuss the physics potential for dark sector searches using future missions such as COSI. We close by summarizing our results in \cref{sec:Conclusions}.

\pagebreak

\section{Neutrino upscattering - Transition Dipole \label{sec:Neutrinos}} 
\begin{figure}
    \centering
    \includegraphics[width = 1\columnwidth]{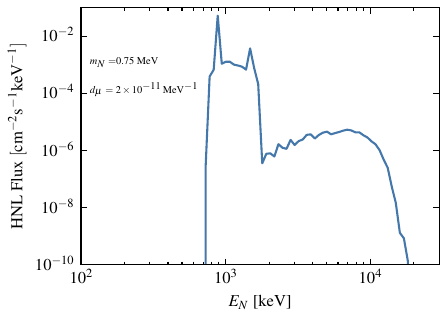}
    \caption{The flux of solar HNLs at Earth (ignoring decays) as calculated through the dipole model Monte Carlo simulation, where $m_N = 0.75$ MeV and $d_{\mu} =2 \times 10^{-11}$ $\mathrm{MeV}^{-1}$.}
    \label{fig:Dipole_HNL_Flux}
\end{figure}
We begin by considering a production mechanism involving the upscattering of solar neutrinos transiting through the Sun, e.g.\ $\nu A \rightarrow {\rm LLP} A$ with $A$ being a nucleus such as hydrogen or helium (see e.g.\ \cite{Plestid:2020ssy,Plestid:2020vqf,Gustafson:2022rsz} for results on neutrino upscattering in the Earth). This mechanism leverages the large solar neutrino flux which is copious in the few-hundred keV region, and extends up to $\sim 15~{\rm MeV}$. Solar neutrinos have a small probability of being absorbed in the SM because of the small charged current scattering cross section at $E_\nu \sim {\rm MeV}$ energies. It is, however, possible to have BSM cross sections that exceed the weak interaction at low energies if neutrinos couple via a transition magnetic dipole moment \cite{Magill:2018jla,Brdar:2020quo}. This can lead to sizable conversion probabilities into an unstable right-handed neutrino, $N$ (also called a heavy neutral lepton or HNL), for neutrinos transiting from the center to the surface of the Sun. As it is unstable, $N$ may decay in flight supplying a broad flux of photons in RHESSI. Similar phenomena may occur in the aftermath of SN 1987A \cite{Brdar:2023tmi,Matheus_Private} leading to tight limits below the supernova floor derived in \cite{Magill:2018jla}.

\begin{figure}
    \centering
    \includegraphics[width = 1\columnwidth]{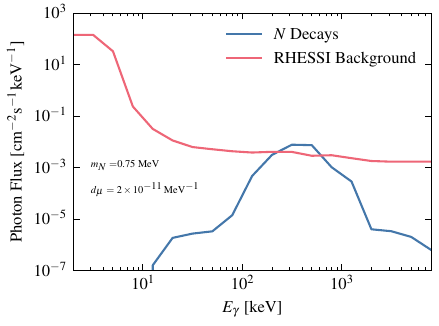}
    \caption{The flux of photons at RHESSI from $N$ decays calculated using a Monte Carlo integration with a $90^{\circ}$ opening angle, compared with the RHESSI background in the front segment. Since the flux from decays exceeds the background, we consider $m_N = 0.75$ MeV, $d_{\mu}= 2 \times 10^{-11}\,\mathrm{MeV}^{-1}$ to be excluded.}
    \label{fig:Dipole_Photon_Flux}
\end{figure}

This ``dipole portal'' can dominate low energy phenomenology since it is a dimension-five operator as compared to the dimension-six contact operator of the weak interaction. Low energy cross sections are then proportional to $d^2$, with $d$ the dipole moment, vs $G_F^2 E^2$ for weak cross sections. Therefore, dipole portal cross sections can be large at low energies while simultaneously avoiding constraints from higher energy experiments (e.g.\ accelerator neutrino experiments and colliders). The effective Lagrangian is given by \cite{Magill:2018jla}
\begin{equation}    
    \label{dip-lag}
    \mathcal{L}_{\rm int} \supset \sum_\alpha d_\alpha F^{\mu\nu}\bar{N} \sigma_{\mu\nu} P_L \nu_\alpha~. 
\end{equation}
Here, $d_{\alpha}$ represents the coupling between $N$ and each of the 3 SM neutrinos. In this work, we consider the cases where $N$ couples to a single flavor. This effective interaction has been studied recently in the context of accelerator, solar, atmospheric, and collider neutrinos as well as in the context of early universe cosmology and constraints from SN 1987A \cite{Gninenko:2009ks,Gninenko:2010pr,McKeen:2010rx,Masip:2011qb,Masip:2012ke,Coloma:2017ppo,Magill:2018jla,Brdar:2020quo,Shoemaker:2018vii,Arguelles:2018mtc,Hostert:2019iia,Fischer:2019fbw,Coloma:2019htx,Schwetz:2020xra,Arina:2020mxo,Shoemaker:2020kji,Abdullahi:2020nyr,Shakeri:2020wvk,Atkinson:2021rnp,Cho:2021yxk,Kim:2021lun,Arguelles:2021dqn,Ismail:2021dyp,Miranda:2021kre,Bolton:2021pey,Jodlowski:2020vhr,Vergani:2021tgc}. 

Unlike the monoenergetic LLP cases discussed later in this paper, the spectrum of $E_{\nu}$ (and hence $E_{N}$ and $E_{\gamma}$) spans several orders of magnitude. For that reason, we implement a Monte Carlo integration to sample neutrino energy, production location, and upscattering location. We also account for flavor transformation between the three SM neutrino flavors during the neutrino propagation (both due to adiabatic conversion and oscillations).

We consider the Sun to be solely comprised of ${}^1$H and ${}^4$He with densities given by the Standard Solar Model \cite{bahcall2004we, Bahcall:2005va,StandardSolar}. Although larger nuclei would have a cross section that scales as $Z^2$ due to coherent effects, the relative abundances of these elements are small so this remains a subdominent effect \cite{Bahcall:2005va}. Different solar models (i.e. \cite{vinyoles2017new}) only differ at the percent-level or less, which is a higher level of precision than considered in this paper. All scattering is calculated to be off free nucleons, ignoring the coherent enhancement due to helium. This only leads to an $\sim 10\%$ change in the bounds, which we will see is a much smaller effect than uncertainty in the detector opening angle/background. The cross section for scattering on a free proton is given by $\dd \sigma_{\mathrm{dip}} = \dd \sigma_1 + \dd \sigma_2$ with, 
\begin{equation}
    \begin{split}
    \frac{\dd\sigma_1}{d E_{r}} =& \alpha (2 d)^2 F_1^2 \bigg ( \frac{1}{E_r} - \frac{1}{E_{\nu}} \\
    &+\frac{m_N^2(E_r-2 E_{\nu} - m_p)}{4 E_{\nu}^2 E_{r} m_{p}} +\frac{m_N^4(E_r-m_p)}{8 E_{\nu}^2 E_{r}^2 m_p^2} \bigg )~,
    \end{split}
\end{equation}
and 
\begin{equation}
    \begin{split}
        \frac{\dd \sigma_2}{dE_r} =& \alpha d^2 \mu_n^2 F_2^2 \bigg [ \frac{2 m_p}{E_{\nu}^2} \big ( (2 E_{\nu}-E_{r})^2 - 2 E_{r} m_{p} \big ) \\
        &+ m_N^2 \frac{E_r - 4 E_{\nu}}{E_{\nu}^2} + \frac{m_N^4}{E_{\nu} E_{r}} \bigg ]~.
    \end{split}
\end{equation}
Here, $F_1$ and $F_2$ are electromagnetic form factors, $\mu_n$ is the magnetic moment of the nucleon in question, $E_r$ is the recoil energy, and $m_p$ is the proton mass \cite{Borah:2020gte, PhysRevC.69.022201}. Since the neutrino energy is much less than the proton mass, the HNL energy $E_{N}$ is nearly identical to the neutrino energy $E_{\nu}$. Thus, the flux of HNLs has similar features to the solar neutrino flux (see \cref{fig:Dipole_HNL_Flux}). Note that for the parameters considered, the flux of HNLs is $\sim$7 orders of magnitude below the solar neutrino flux, so we do not expect upscattering to have a noticeable effect on solar neutrino detection. The downscattering rate of HNLs will similarly be small, so we do not consider any scattering after production.

The HNL has decay channels $N \rightarrow \nu_{\alpha} \gamma$. We consider the $\nu$ to be massless, and the decays to be isotropic in the rest frame of the HNL.\footnote{In complete generality the HNL may have some angular correlation with its polarization, but this depends on the details of the model e.g.\ Dirac vs.\ Majorana neutrinos \cite{Balantekin:2018ukw} and we neglect this in what follows.} Using the relativistic quantities $\gamma = E_N/m_N$ and $\beta = \sqrt{1 - m_{N}^2/E_{N}^2}$, the decay length is calculated as
\begin{equation}
    \lambda = \frac{4 \pi}{d_{\alpha}^2 m_N^3} \gamma \beta .
    \label{eq:Decay_Length}
\end{equation}

The Monte Carlo simulation samples locations for $N$ decays along with the energy and direction of the decay photon. This is used to calculate the resulting photon flux with respect to energy and angle observed by RHESSI. We consider opening angles for HNL decay photons of $1^{\circ}$ and $90^{\circ}$, where we reject all photons arriving at larger angles. The true opening angle is expected to be an energy-dependent value between these two angles, but a full analysis is beyond the scope of this paper. The background flux observed by RHESSI is calculated by using the reported number of counts and effective area of the front segment (ignoring narrow peaks) \cite{smith2003rhessi}. We reject a parameter point if the flux from $N$ decays exceeds the observed flux at any energy (see \cref{fig:Dipole_Photon_Flux}). RHESSI has a good energy resolution (3 keV FWHM at energies below 1 MeV, and similar resolutions at higher energies) so we expect the effect to be noticable even if the BSM photon flux only exceeds the background flux for a small range of energies.
\begin{figure}
    \centering
    \includegraphics[width = 1\columnwidth]{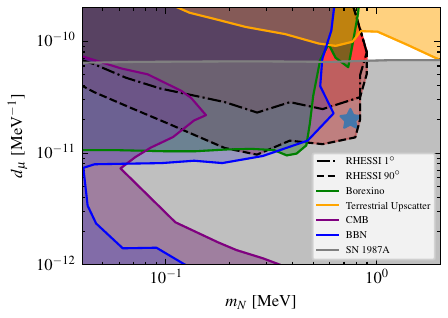}
    \caption{Excluded parameter space for a muon neutrino transition dipole moment. Along with our bounds, we show 90\% CL exclusions from Borexino $e-\nu$ scattering \cite{Brdar:2020quo,borexino2018comprehensive}, terrestrial solar neutrino upscattering \cite{Plestid:2020vqf}, Supernova 1987A \cite{Magill:2018jla}, big-bang nucleosynthesis and the cosmic microwave background \cite{Brdar:2020quo}. For RHESSI excluded parameter space, we include exclusions from taking a $1^{\circ}$ opening angle and a $90^{\circ}$ opening angle for photons from HNL decays. For both opening angles, we use the same background. The star represents the parameter point show in \cref{fig:Dipole_HNL_Flux} and \cref{fig:Dipole_Photon_Flux}.}
    \label{fig:Dipole_Exclusion}
\end{figure}

\begin{figure}
    \centering
    \includegraphics[width = 1\columnwidth]{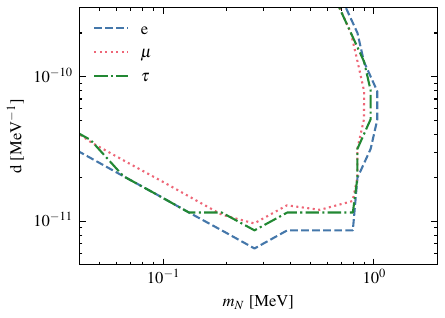}
    \caption{Excluded parameter space of a transition dipole moment for each of the three active neutrinos. We see that the constraints all take a similar form, only varying by $\mathcal{O}(1)$ factors.}
    \label{fig:Dip_Flavor_Compare}
\end{figure}

Our resulting exclusion curves from the RHESSI data are shown in \cref{fig:Dipole_Exclusion} for a muon neutrino dipole coupling. We find that RHESSI data can offer a complimentary (and direct) probe of regions of parameter space that are already probed by SN-1987A. Constraints are strongest in the low mass region (sub-MeV), and this may also be probed using coherent elastic neutrino nucleus scattering. We see that the exclusions for the three neutrino flavors all have similar values in \cref{fig:Dip_Flavor_Compare}. Although the solar neutrino energy extends to $\sim$ 15 MeV, the bounds end near $m_N = 1$ MeV. Beyond this mass, the decay length becomes significantly smaller than the radius of the Sun (as can be seen in \cref{eq:Decay_Length}), so HNLs are unable to escape before decaying.

\section{Heavy solar axions \label{sec:Axions}}

Another production mechanism is solar axions with energies in excess of $E_a\gtrsim 500$ keV. These energies are too high to allow for thermal production (except for in exponentially suppressed tails), and so the background photon fluxes are much smaller than for typically considered keV solar axion searches. The study of MeV-scale solar axions has a long history, and they have been searched for in terrestrial experiments such as Borexino and CAST \cite{CAST:2009klq,Borexino:2012guz}. As we discuss below, satellite measurements of the Quiet Sun provide a complimentary probe that excels for decay lengths that are short relative to the Earth-Sun distance.

It is worth highlighting recent work on model building for axions with an extended matter content \cite{Fukuda:2015ana,Agrawal:2017ksf,Hook:2019qoh,Dunsky:2023ucb}. These models are motivated by the axion quality problem and seek to protect the axion against Planck suppressed corrections. The simplest mechanism to achieve this is to simply break the canonical relation $f_a m_a \approx f_\pi m_\pi$ and to allow for $m_a$ to be ``heavy'' relative to predictions of conventional (i.e.\ DSVZ \cite{Dine:1981rt,Zhitnitsky:1980tq} or KSVZ \cite{Kim:1979if,Shifman:1979if}) axion models. It is interesting to note that these independent model building considerations often push the mass and couplings of the axion into regions of parameter space that are well suited for solar axion detection; we will comment on this in great detail below. For instance, following the benchmark scenarios presented in \cite{Dunsky:2023ucb} one finds that masses in the $\sim 10$ MeV regime with axion decay constants $f_a \sim 10^{-5}~{\rm GeV}^{-1}$ fall squarely within the ``natural'' window of parameter space whilst simultaneously predicting a sizeable coupling to nucleons and a decay length that is a few times longer than the radius of the Sun. For slightly lighter axions, solar production and detection is a useful complimentary probe. 

In this section we will parameterize constraints in terms of low energy constants of the effective theory describing axion interactions with nucleons and photons. This may be parameterized by the Lagrangian
\begin{equation}
    \mathcal{L}_{\rm int} \subset g_{a\gamma\gamma} a F^{\mu\nu} \tilde{F}_{\mu\nu}  + \frac{g_{3aN}}{m_N} (\partial_\mu a) \overline{N} \gamma^\mu \tau_3  N ~,
\end{equation}
We focus on the isovector coupling because of the $M1$ transition relevant for phenomenology in the Sun. We will allow $g_{a\gamma \gamma}$, the coupling controlling the rate of $a\rightarrow \gamma \gamma$, to vary independently of the isovector coupling of axions to nucleons, $g_{3aN}$. In a UV-completion these parameters will be tightly correlated and expressible in terms of the axion decay constant $f_a$. A reasonable order of magnitude estimate is that $g_{3 aN} \sim m_N/f_A$ and $g_{a\gamma \gamma} \sim \frac{\alpha}{4\pi} \frac{1}{f_a}$, however details are model dependent and we do not discuss them further.

The primary production mechanism for heavy solar axions is the $p~d \rightarrow \!~^{3}{\rm He} ~ \gamma$ reaction which takes place in the solar $pp$ chain. Other mechanisms are energetically allowed, such as $M1$ transitions in the CNO chain \cite{Massarczyk:2021dje}, and $e^+e^-$ annihilation from $^{8}{\rm B}$ neutrinos in the solar interior, however, we find that the production rates are sufficiently small so as to be uninteresting. 

The flux of axions (prior to decay) can be related to the flux of $pp$ neutrinos, and depends on the isovector coupling of axions to nucleons $g_{3aN}$ \cite{Donnelly:1978ty}. The axions must first escape the Sun and then decay before reaching Earth. The escape probability depends both on axion absorption and decay processes. Putting all of this together and setting ${\rm BR}_{a \gamma\gamma}=1$, we arrive at the flux of axions arriving at a detector orbiting the Earth, 
\begin{equation}
    \frac{\Phi_\gamma}{\Phi_{\nu}^{(pp)} } = 0.54 |g_{3aN}|^2 \qty[\frac{p_a}{p_\gamma}]^3  \qty[ \e^{-R_\odot/\ell_{\rm abs}} - \e^{-d_{\odot}/\ell_{\rm dec}}] ~, 
\end{equation}
where $\ell_{\rm abs}^{-1} = \ell_{\rm MFP}^{-1} + \ell_{\rm dec}^{-1}$ with $\ell_{\rm MFP}^{-1}$ the averaged mean free path in the Sun and $\ell_{\rm dec}$ the axion decay length. The coupling $g_{3aN}$ is the isovector coupling strength of the axion to nucleons, and $p_a/p_\gamma$ is the ratio of three-momenta between an axion and photon emitted with $E=5.49~{\rm MeV}$. The $pp$ neutrino flux is given by 
$\Phi_{\nu}^{(pp)}=6 \times 10^{10}~{\rm cm}^{-2}{\rm s}^{-1}$. We account for axion-absorption, Primakoff scattering, and axion electron scattering in our calculation of $\ell_{\rm MFP}^{-1}$. The resulting photon flux will be constant in energy over the kinematically allowed photon energies.

\begin{figure}
    \centering
    \includegraphics[width = 1\columnwidth]{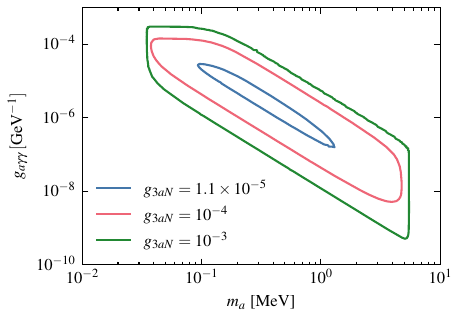}
    \caption{Contours of $g_{3aN}$ for which the solar axion flux of photons would overwhelm the the RHESSI background measurements for the front segments. Sensitivity is exhausted for $g_{3aN} \sim 1\times 10^{-5}$ however further reach can be obtained with better data and/or a more sophisticated analysis. To compare with constraints from supernovae \cite{PhysRevLett.128.221103} and rare kaon decays \cite{NA62:2021zjw,E787:2004ovg,BNL-E949:2009dza} one can use a naive estimate of $g_{3aN} = m_N / f_a$ and $g_{a\gamma \gamma} = \frac{\alpha}{4\pi} \frac{1}{f_a}$, and note that $3\times 10^4 ~{\rm GeV} \lesssim f_a \lesssim 3\times 10^5 ~{\rm GeV}$ is allowed by the above constraints (see e.g.\ the ``conservative'' curves in Fig.\ 9 of Ref.\ \cite{Dunsky:2023ucb}). As an illustration taking $m_a=1~{\rm MeV}$ and $f_a=3 \times 10^4~{\rm GeV}$, one finds $g_{a\gamma \gamma} \sim 2\times 10^{-8}$ and $g_{3aN} \sim 3\times 10^{-5}$.}
    \label{fig:axion_exclusion}
\end{figure}

Our results are shown in \cref{fig:axion_exclusion}. We note that our exclusions depend on the axion nucleon coupling, captured by $g_{3aN}$, and the decay constant $g_{a\gamma \gamma}$. If $g_{a\gamma\gamma}$ vanishes at some scale $\mu=\mu_0$, but $g_{aee}\neq 0$ then an effective $g_{a\gamma\gamma}\sim (\alpha/4\pi) g_{aee}/m_e$ will be generated via a 1-loop triangle diagram, and in this way one can re-cast our limits\footnote{This requires accounting for the branching ratio to photons, as well as adjusting the decay length.} in terms of those on $g_{aee}$. We do not include exclusions from SN1987 typically plotted in the $m_a-g_{a\gamma\gamma}$ plane because the values of $g_{3aN}$ that are required to produce a sufficient axion flux in the Sun lead to axion trapping within a core-collapse supernova \cite{Chang:2018rso}.\footnote{This occurs because axion-nucleon scattering leads to mean free paths much shorter than the typical size of a supernova, trapping the axions.} This is an important distinction between the hadronically coupled axion models we considered here vs.\ an axion like particle which couples exclusively to photons (see e.g.\ \cite{Jaeckel:2015jla}). The solar axion constraints we discuss here are therefore complimentary to supernova cooling ones. If the axion nucleon coupling, $g_{aN}$, is large enough to evade SN-1987 bounds via self trapping then it is also large enough to be probed with RHESSI data. Low energy supernovae observations have been used to place constraints on axions which decay in-flight and deposit energy to the ejecta \cite{PhysRevLett.128.221103}. Additionally, axions produced in neutron star mergers have been constrained using X-ray observation \cite{diamond2023multimessenger}. These constraints also disappear in the strong coupling regime, i.e.\ for $f_a \lesssim 3 \times 10^5~{\rm GeV}$, and are complimentary to ours. Constraints from NA62 \cite{NA62:2021zjw}, E787 \cite{E787:2004ovg}, and E949 \cite{BNL-E949:2009dza} are subject to $O(m_K^4/m_\rho^4)$ hadronic uncertainties in the prediction of $K\rightarrow a \pi$ \cite{Bauer:2021wjo,Dunsky:2023ucb}. These constraints require $f_a \gtrsim 3\times 10^{4}$. Finally, our constraints on $g_{a\gamma\gamma}$ lie above the ceiling of searches performed with the Borexino collaboration \cite{Borexino:2012guz} because we are sensitive to decay lengths much shorter than $d_{\odot}$. This is demonstrative of the way in which constraints from solar axion may compliment existing search techniques using accelerator based experiments, underground detectors, and astrophysical constraints. 

Constraints from big bang nucleosynthesis (BBN) will generically apply both because the axions we consider have lifetimes in the vicinity of a few seconds, and because the same reaction, $p~d\rightarrow\!~^{3}{\rm He}~\gamma$, is a key driver of BBN. In the absence of any additional dark sector decay modes, measurements of $N_{\rm eff}$ will generically exclude axions with masses below 5 MeV or so. These constrains can be alleviated if the dark sector contains additional degrees of freedom see e.g.\ \cite{Dunsky:2023ucb}. Searches for gamma rays from the Quiet Sun offer a complimentary direct probe of axion (or other light particle) production that is independent of early universe cosmology. 

We consider a $90^{\circ}$ opening angle for our signal, meaning all decays between the Sun's surface and Earth's orbit contribute. The monoenergetic nature of the axion means the photon flux is constant in energy (see \cref{sec:Model_Ind} for more details on monoenergetic production). We demand that this flux exceed $1.8 \times 10^{-3} \mathrm{s}^{-1} \mathrm{cm}^{-2} \mathrm{keV}^{-1}$ for photon energies above 1 MeV, so that this flux is above the observed RHESSI background flux in the front segments.

\section{Model-Independent Searches \label{sec:Model_Ind}}
Let us now consider a model-independent production of LLPs (here called $\phi$) which decay via $\phi \rightarrow \gamma \gamma$. In this simplified model, we consider all production to occur at the solar center, and $\phi$ only interacts with SM physics through its decay, so we ignore any possible scattering or absorption. We also assume there is no preferential direction for decay in the rest frame of $\phi$, so the flux of photons is a uniform distribution between $E_{\gamma, \mathrm{min}}$ and $E_{\gamma, \mathrm{max}}$ where $E_{\gamma, \mathrm{max/min}} = 1/2 \times  ( E_{\phi} \pm \sqrt{E_{\phi}^2 - m_{\phi}^2} )$. Inverting this equation, we find $E_{\phi} \geq E_{\gamma} + m_\phi^2/(4E_{\gamma})$ (we will call this lowest energy $E_{\phi,\mathrm{min}}$). Therefore, if we know the rate of production $R_{\phi}$ and decay length $\lambda$ as a function of $E_{\phi}$, then we can determine the BSM flux of photons at Earth.
\begin{equation}
    \begin{split}
    \frac{\dd  \Phi_{\gamma}}{\dd E_{\gamma}} =& \frac{2}{4\pi d_{\odot}^2} \times \\
    &\int^{\infty}_{E_{\phi,\mathrm{min}}} \dd E_{\phi}  \frac{e^{-R_{\odot}/\lambda(E_{\phi})} - \e^{-d_{\odot}/\lambda(E_{\phi})}}{\sqrt{E_{\phi}^2-m_{\phi}^2}} \frac{\dd R_{\phi}}{\dd  E_{\phi}}
    \end{split}
    \label{eq:PhiToPhotonIntegral}
\end{equation}

One particularly well motivated morphology is where $\phi$ has a mono-energetic production spectrum. We have already seen an example of this in the case of monoenergetic axions considered in \cref{sec:Axions}. This would also occur if $\phi$ is produced via a 2-body decay $\chi \rightarrow \phi X$ or via annihilation $\chi \chi \rightarrow \phi X$ for $v_\chi \ll 1$ and $\chi$ and X are some generic particles. Performing the integral in \cref{eq:PhiToPhotonIntegral} with a delta-function distribution leads to a flux of photons that is constant in energy between $E_{\gamma, \mathrm{min}}$ and $E_{\gamma, \mathrm{max}}$. 

\begin{figure}
    \centering
    \includegraphics[width = 1\columnwidth]{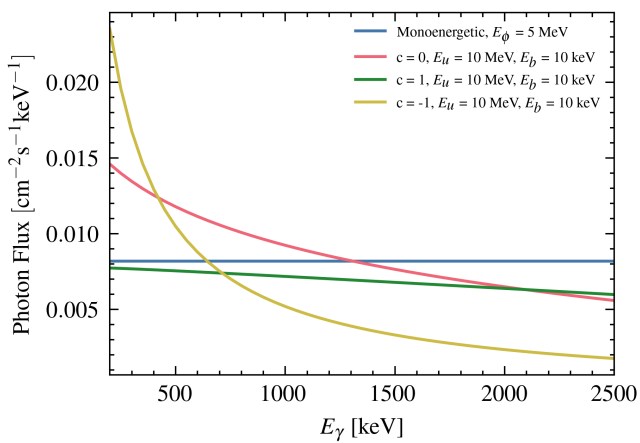}
    \caption{Comparison of photon fluxes for different $\phi$ production scenarios. The fluxes are normalized so that the total production rate is $N_{\phi} = 10^{28} s^{-1}$, and the decay length is $\lambda = 10 R_{\odot}$ at $E_{\phi} = 1 \mathrm{MeV}$. We consider $m_{\phi}$ to be negligibly small.}
    \label{fig:LLP_Flux_Comparison}
\end{figure}

\begin{figure*}[t]
    \subfloat[Mono-energetic $\phi$ production \label{fig:phi_mono}]{%
      \includegraphics[width=0.475\textwidth]{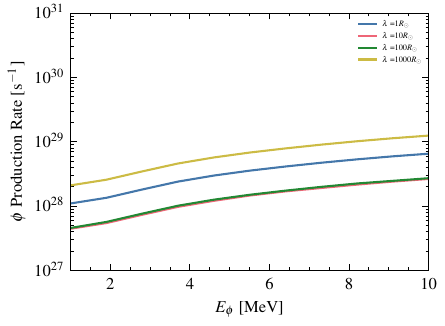}
    }
    \hfill
    \subfloat[$\phi$ production uniform in energy (c = 0) \label{fig:phi_uni}]{%
      \includegraphics[width=0.475\textwidth]{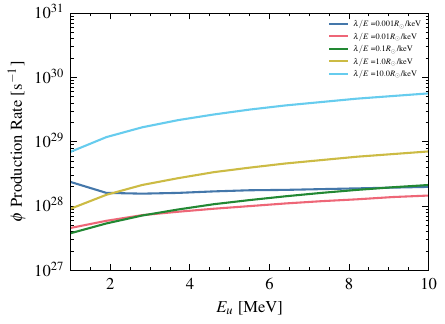}
    }

    \subfloat[$\phi$ production with a linear dependence on energy (c = 1) \label{fig:phi_lin}]{%
      \includegraphics[width=0.475\textwidth]{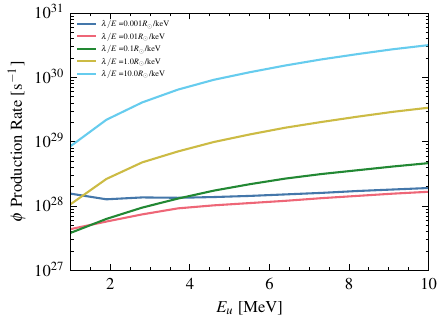}
      }
    \hfill
    \subfloat[$\phi$ production inversely proportional to energy (c = -1). \label{fig:phi_neg_1}]{ \includegraphics[width = 0.475\textwidth]{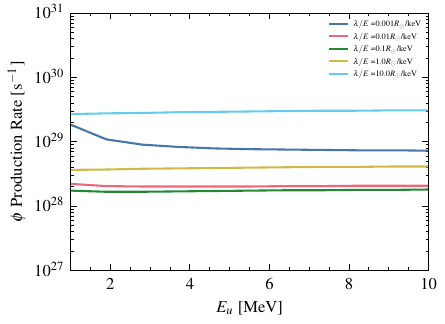}}

        \caption{Exclusion of $\phi$ production for some of the special cases considered. Production rates above the lines are excluded. In all cases, the mass is considered negligible, and there is no production below 10 keV ($E_{b} = 10\, \mathrm{keV}$). \label{LLP_Exclusion_Plot}}
  \end{figure*}

Remaining more agnostic to the source of $\phi$ production, we may consider a power-law distribution with respect to energy for $E_{b} \leq E_{\phi} \leq E_{u}$
\begin{equation}
    \frac{\dd  R_{\phi}}{ \dd E_{\phi}} \bigg \vert_{\mathrm{power}} = R_{c} \times E_{\phi}^c ~\Theta(E_{u} - E_{\phi}) \Theta(E_{\phi} - E_{b})~.
    \label{eq:Power_Phi}
\end{equation}
For $m_{\phi} \ll E_{\gamma}, E_{\phi}$ the photon flux is calculable in closed form, 
\begin{equation}
    \begin{split}
    &\frac{\dd \Phi_{\gamma}}{\dd E_{\gamma}} \bigg \vert_{\mathrm{power}} = \frac{2 R_c}{4 \pi d_{\odot}^2} \\
    &\times \bigg [\bigg (\frac{R_{\odot}\Tilde{E}}{\Tilde{\lambda}}\bigg )^c \bigg (\Gamma \big (-c, \frac{R_{\odot} \Tilde{E}}{\Tilde{\lambda} E_u} \big ) - \Gamma \big (-c, \frac{R_{\odot} \Tilde{E}}{\Tilde{\lambda} E_l} \big) \bigg) \\
    &- \bigg (\frac{d_{\odot}\Tilde{E}}{\Tilde{\lambda}} \bigg)^c \bigg (\Gamma \big(-c, \frac{d_{\odot} \Tilde{E}}{\Tilde{\lambda} E_u} \big) - \Gamma \big (-c, \frac{d_{\odot} \Tilde{E}}{\Tilde{\lambda} E_l} \big) \bigg) \bigg]~, 
    \end{split}
\end{equation}
with $\Gamma(a, x)$ the incomplete gamma function, $E_l = \max\{E_b,E_{\phi,\mathrm{min}}\}$, and $\Tilde{\lambda}$ is the decay length at characteristic energy $\Tilde{E}$. We normalize to the total rate of $\phi$ produced, $N_{\phi}$. For the mono-energetic case, we have $N_{\phi} = R_{\phi}$, while for the power-law production, we have
\begin{equation}
   R_{c} = \begin{cases}
         \frac{N_{\phi} (c+1)}{E_u^{c+1}-E_b^{c+1}} &\mathrm{for}~~~c \neq -1~, \\
        \frac{N_{\phi}}{\log(E_u/E_b)} & \mathrm{for}~~~ c = -1~.
    \end{cases}
\end{equation}
Constraints on the number of $\phi$ produced per second in the Sun are shown in \cref{LLP_Exclusion_Plot}. Constraints are set as described at the end of \cref{sec:Neutrinos,sec:Axions}.

\section{Future prospects \label{Future} }

In the above discussion, we have found that re-purposing existing RHESSI data is able to provide interesting constraints on light dark sectors with MeV scale LLPs. Our analysis should be viewed as a proof of principle and certainly underestimates the sensitivity of experiments like RHESSI to new physics models. The major limitations in our analysis are a lack of reliable peak-subtracted spectra and the ability to suppress backgrounds (see \cite{DeRocco:2022jyq,frerick2023view} for recent work in the keV regime for more sophisticated statistical analyses). For example, much of the background for RHESSI comes not from solar activity but rather from cosmic ray interactions with the Earth's atmosphere i.e.\ the radiation comes from the rear rather than forward field of view. Much of this background can presumably be suppressed (or perhaps eliminated) with a future instrument, especially if a dedicated search is performed. As our current analysis is systematically limited, an experiment with 10\% of RHESSI's background (or with the same background but modeled to a 10\% uncertainty) would be 10 times more sensitive to a BSM flux of photons.  In what follows we sketch potential improvements using a near-term MeV telescope. For concreteness we will anchor our discussion around the COSI satellite.\footnote{We thank Albert Shih for pointing out the COSI mission to us.} 

RHESSI operated with minimal shielding to minimize weight. This made the instrument an effectively ``all sky'' observation with a high level of cosmic ray background activity. In contrast COSI will operate with active shielding, and its further use of Compton kinematic discrimination offers further background reduction \cite{Boggs:2000zk}. Moreover, ongoing work to better understand gamma ray emission from the Quiet Sun will further improve on irreducible backgrounds \cite{
Orlando:2020ezh,Orlando:2022xsm}.

Other strategies that could be pursued with a future instrument are to go beyond the rate-only analysis presented above. For example, COSI will have 25\% sky coverage and excellent angular resolution. One could image the MeV photon flux differential in both energy and angular position. Depending on the lifetime of LLPs a ``halo'' of photons could be searched for outside the solar corona. The shape of the photon distribution will be model dependent, but can be computed using the Monte Carlo simulations outline above. Similarly, taking advantage of COSI's large field of view, other local planetary systems could be used to search for LLPs. This was suggested recently in the context of Jupiter where the capture of light dark matter is better motivated \cite{Leane:2021tjj,Acevedo:2023owd}.

Finally, let us comment on a second channel of interest: ${\rm LLP} \rightarrow e^+e^-$. This may occur for a dark vector which dominantly decays via $V\rightarrow e^+e^-$, and has recently been considered (in the context of large volume underground detectors) for the same $p~d\rightarrow \! ~^{3}{\rm He}~\gamma$ reaction considered here \cite{DEramo:2023buu}. A search for electrons and/or positron would require accurate modeling for propagation through magnetic fields in the vicinity of the Earth.

\vfill
\pagebreak

\section{Conclusions and Outlook \label{sec:Conclusions}}

We have discussed simple particle physics models that predict an MeV flux of photons produced by the Sun. The generic requirement is the existence of some LLP which can efficiently transport energy from the interior (fueled by nuclear reactions) to beyond the Sun's surface. Provided the LLP has a sizeable branching ratio to final states including at least one photon e.g.\  $\gamma \gamma$, $\nu \gamma$, and/or $e^+e^-\gamma$ final states, one can search for energetic gamma rays emanating from the Quiet Sun. 

We find that constraints from existing data from RHESSI, with a very conservative analysis strategy, can probe small pockets of untouched parameter space for both MeV scale axions and a neutrino dipole portal. In both cases, the RHESSI analysis provides complimentary coverage to existing search strategies (including cosmological probes such as BBN). 

Our major motivation is a simple proof of principle that MeV-scale LLPs with decay lengths larger than the radius of the Sun can be efficiently searched for using solar telescopes. The analysis presented here is conservative and fairly crude; we define exclusions by the condition that the BSM signal prediction exceeds the {\it total signal} observed in any energy window by RHESSI. Constraints and/or discovery potential could be substantially improved with a better understanding of instrument backgrounds and more sophisticated analysis techniques. For example, one could make use of angular profiles of incident photons to search for new physics, as an LLP flux will produce a photon flux outside the stellar corona with a predictable angular shape/morphology. We encourage future missions with MeV scale instrumentation below the cut-off of Fermi-Lat, such as COSI \cite{COSI_2023,Tomsick:2021wed}, to consider searches for BSM particles, with the Sun being a well-motivated engine for MeV-scale LLPs.

\section*{Acknowledgments} 
 This work benefited from feedback at the Simon's Center for Geometry and Physics, and RP would like to specifically thank Simon Knapen, Rebecca Leanne, and Jessie Shelton for useful discussions. We thank Albert Shih for helpful discussions regarding the RHESSI instrument. We benefited from feedback on this manuscript from Rebccea Leanne and Elena Pinetti. 

RP is supported by the U.S. Department of Energy, Office of Science, Office of High Energy Physics, under Award Number DE-SC0011632 and by the Walter Burke Institute for Theoretical Physics. RP is supported by the Neutrino Theory Network under Award Number DEAC02-07CHI11359, the U.S. Department of Energy, Office of Science, Office of High Energy Physics, under Award Number DE-SC0011632, and by the Walter Burke Institute for Theoretical Physics. IMS and RAG are supported by the U.S. Department of Energy Office of Science, Office of High Energy Physics, under Award Number DE-SC0020262.

\vfill
\pagebreak

\appendix

\section{Inefficient production mechanisms\label{app:other_mechanisms}}
In this section we discuss production mechanisms which we have found to be too inefficient to allow for detection prospects with our RHESSI analysis.

\subsection{Mass-Mixing portal for HNLs} 
\begin{figure}
    \centering
    \includegraphics[width = 1\columnwidth]{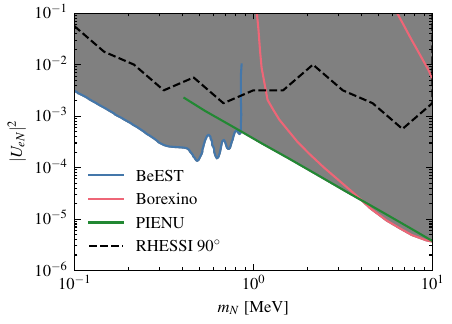}
    \caption{Excluded parameter space for HNLs in the mass mixing model. The dashed line shows the RHESSI exclusion, while shaded regions come from BeEST  \cite{friedrich2021limits}, PIENU \cite{bryman2019improved}, and Borexino  \cite{bellini2013new}.}
    \label{fig:Mass_Mix_Exclusion}
\end{figure}

Another BSM model involving HNLs has $N$ couple directly to active neutrinos through added elements in the PMNS matrix\cite{Atkinson:2021rnp,Plestid:2020ssy,friedrich2021limits,bryman2019improved,bellini2013new, Arguelles:2019ziu,Liventsev:2013zz,Asaka:2005pn,Asaka:2005an,Atre:2009rg,Johnson:1997cj,Levy:2018dns,Formaggio:1998zn,Gorbunov:2007ak,Drewes:2013gca,Bondarenko:2018ptm,Boyarsky:2009ix,Berryman:2019dme,Ballett:2019bgd,Gelmini:2019deq,Orloff:2002de,Boiarska:2021yho,Sabti:2020yrt,PhysRevLett.127.121801}. Active neutrinos contain a small admixture of the HNLs along with the three known mass states, 
 \begin{equation}
    \nu_\alpha = U_{\alpha N} N + \sum_{i=1}^3 U_{\alpha i}\nu_i~,
 \end{equation}
where $U_{\alpha N}$ represents the coupling of HNLs to active neutrinos. Since the Sun only has nuclear reactions that produce electron neutrinos, our constraint is on $U_{e N}$. The $N$ flux from upscattering is subdominant by orders of magnitude to that from direct production. Therefore, the flux is given by rescaling the neutrino flux
\begin{equation}
    \Phi_{N} = |U_{N e}|^2 \Phi_{\nu} \sqrt{1 - m_N^2/E_N^2} .
    \label{eq:Relative_HNL_Flux}
\end{equation}
For the masses considered here, there are only three decay channels; {\it i)} $N \rightarrow 3 \nu$, {\it ii)} $N \rightarrow \nu \gamma$, and {\it iii)} $N \rightarrow \nu e^{+} e^{-}$. As with other production mechanisms, we only consider signals from photons. The geometry of this decay (into a massless neutrino and photon) is identical to the case of the dipole portal. The decay rate for each of the processes follows the general form
\begin{equation}
    \Gamma_{N \rightarrow SM} \propto G_F^2 |U_{e N}|^2 m_N^5~,
\end{equation}
which has the steep power-law dependence on mass typical of weak decays. We find that since decay lengths are always long enough to fall outside the range given in \cref{eq:dec_range} that sensitivity from RHESSI is subdominant to searches at Borexino (which benefits from a large detector volume) and from direct laboratory searches (see \cref{fig:Mass_Mix_Exclusion}). 

\subsection{Captured dark matter in the Sun}
If heavy dark matter, $\chi$, has interactions beyond gravity, it may scatter within large celestial bodies and become gravitationally captured. The Sun, being by far the most massive object in the solar system, is a strong candidate in searching for the signals from captured $\chi$ \cite{Press:1985ug,Gould:1987ju,gould1987resonant,garani2017dark,frandsen2010asymmetric,PhysRevD.82.115012,bell2011enhanced,PhysRevD.93.115036, PhysRevD.95.075001,PhysRevD.95.123016,PhysRevD.96.063010,albert2018constraints,Nisa:2019mpb,niblaeus2019neutrinos, PhysRevD.101.022002,Serini:2019u2,PhysRevD.102.022003,Bell:2021pyy}.

For the case of symmetric dark matter with a long-lived particle mediator, there is the interaction $\chi \Bar{\chi} \rightarrow \mathrm{LLPs}$. The energies of these final observable particles are $\mathcal O (m_{\chi})$. However, as discussed in \cite{garani2017dark}, for thermal relic annihilation cross sections, short range interactions with SM, and $m_{\chi}$ below a few GeV, most of the $\chi$ evaporates from the Sun before annihilating. Even Jupiter, which has a cooler core than the Sun, would have evaporation be the dominant effect for $m_{\chi} \leq 0.7 \mathrm{GeV}$ \cite{garani2022evaporation}, far above the energy sensitivity of RHESSI. We note that in the presence of long-range $\chi - \mathrm{SM}$ interactions, evaporation may be suppressed \cite{Leane:2021tjj,Acevedo:2023owd}. However, this is a model-dependent scenario, and is not considered in this work.

We also considered the case of asymmetric dark matter with self-interactions via a scalar $\phi$ with a Yukawa like interaction $\mathcal{L} \subset \bar{\chi} \chi \phi$ . As there is no annihilation, in the absence of evaporation, the $\chi$ population grows indefinitely. Virialized dark matter passing through the Sun can scatter on the trapped overdensity and produce LLPs via the bremsstrahlung-like reaction $\chi \chi \rightarrow \chi \chi \phi$. In order to produce MeV gamma rays, we require heavy dark matter, $m_\chi \gtrsim  1~{\rm TeV}$, such that there is sufficient available kinetic energy $m_\chi v_\chi^2 \gtrsim 1~{\rm MeV}$.\footnote{Dark matter nucleon scattering cannot induce MeV bremsstrahlung (i.e.\ via $\chi N \rightarrow \chi N \phi$), because the available kinetic energy is set by $m_N  v_\chi^2 \sim 1~{\rm keV} \times (v_\chi/10^{-3})^2$. This is most easily seen in the rest frame of the heavy dark matter.} In order to produce a sufficiently large flux of LLPs, we require a sizeable $\chi \chi \rightarrow \chi \chi$ cross section. This can only be achieved with a light mediator for TeV scale (or heavier) dark matter. The cross section relies on a small momentum transfers. Non-relativistic kinematics, however, demand a parametrically larger momentum transfer in the bremsstrahlung like reaction than for elastic scattering. For example, demanding $E_\phi\sim \mathcal{O}(\mathrm{MeV})$ bremsstrahlung, requires a momentum transfer on the order of $\Delta p^2 \sim  m_{\chi}E_{\phi} \sim (1~{\rm GeV})^2$. Due to this kinematic supression, we find that RHESSI is incapable of setting competitive limits even with the most generous/optimistic model building choices to maximize the bremsstrahlung like cross section. 
 
\bibliography{biblio.bib}

\end{document}